\documentclass[aps,prb,twocolumn,superscriptaddress]{revtex4-1}
\usepackage{graphicx}% Include figure files
\usepackage{dcolumn}% Align table columns on decimal point
\usepackage{bm}% bold math
\usepackage{gensymb} % general symbols

\begin{document}

\title{Crystal growth and annealing study of fragile, non-bulk superconductivity in YFe$_2$Ge$_2$}

\author{H. Kim}
\email[]{hyunsoo@iastate.edu}
%\homepage[]{Your web page}
%\thanks{}
\affiliation{Department of Physics and Astronomy, Iowa State University, Ames, IA 50011}

\author{S. Ran}
\affiliation{Department of Physics and Astronomy, Iowa State University, Ames, IA 50011}
\affiliation{Ames Laboratory, Iowa State University, Ames, IA 50011}

\author{E. D. Mun}
\affiliation{Department of Physics and Astronomy, Iowa State University, Ames, IA 50011}

\author{H. Hodovanets}
\affiliation{Department of Physics and Astronomy, Iowa State University, Ames, IA 50011}
\affiliation{Ames Laboratory, Iowa State University, Ames, IA 50011}

\author{M. A. Tanatar}
\affiliation{Department of Physics and Astronomy, Iowa State University, Ames, IA 50011}
\affiliation{Ames Laboratory, Iowa State University, Ames, IA 50011}

\author{R. Prozorov}
\affiliation{Department of Physics and Astronomy, Iowa State University, Ames, IA 50011}
\affiliation{Ames Laboratory, Iowa State University, Ames, IA 50011}

\author{S. L. Bud'ko}
\affiliation{Department of Physics and Astronomy, Iowa State University, Ames, IA 50011}
\affiliation{Ames Laboratory, Iowa State University, Ames, IA 50011}

\author{P. C. Canfield}
\affiliation{Department of Physics and Astronomy, Iowa State University, Ames, IA 50011}
\affiliation{Ames Laboratory, Iowa State University, Ames, IA 50011}

\date{\today}

\begin{abstract}
We investigated the occurrence and nature of superconductivity in single crystals of YFe$_2$Ge$_2$ grown out of Sn flux by employing x-ray diffraction, electrical resistivity, and specific heat measurements. We found that the residual resistivity ratio (RRR) of single crystals can be greatly improved, reaching as high as $\sim$60, by decanting the crystals from the molten Sn at $\sim$350\celsius~and/or by annealing at temperatures between 550\celsius~and 600\celsius. We found that samples with RRR $\gtrsim$ 34 showed resistive signatures of superconductivity with the onset of the superconducting transition $T_c\approx1.4$ K. RRR values vary between 35 and 65 with, on average, no systematic change in $T_c$ value, indicating that systematic changes in RRR do not lead to comparable changes in $T_c$. Specific heat measurements on samples that showed clear resistive signatures of a superconducting transition did not show any signature of a superconducting phase transition, which suggests that the superconductivity observed in this compound is either some sort of filamentary, strain stabilized superconductivity associated with small amounts of stressed YFe$_2$Ge$_2$ (perhaps at twin boundaries or dislocations) or is a second crystallographic phase present at levels below detection capability of conventional powder x-ray techniques.
\end{abstract}


%\keywords{}

\maketitle

\newcommand{\Yott}{YFe$_2$Ge$_2$}
\newcommand{\Luott}{LuFe$_2$Ge$_2$}

%%%%%%%%%%%%%%%
%%%%%%%%%%%%%%%
%%%%%%%%%%%%%%%
\section{Introduction}

Unconventional superconductivity is frequently found in the vicinity of magnetism.\cite{Mathur1998} Often, when this happens, spin fluctuations are thought to be the pairing glue that leads to superconductivity. Recently, the Fe-based superconductors\cite{Stewart2011} were discovered where the order parameter changes its sign at different parts of the Fermi surface.\cite{Mazin2008} For this class of compounds, superconductivity appears when the antiferromagnetic ordering is adequately suppressed by chemical doping or mechanical pressure. Whereas the AEFe$_2$As$_2$ (AE=Ba,Sr,Ca) based superconductors are quite robust and do not manifest pathologically strong suppression of superconducting transition temperature, $T_c$, by small scattering, some non-$s$-wave superconductors are very susceptible to complete $T_c$ suppression by scattering associated with relatively low residual resistivity values. For instance, the critical residual resistivity values (above which no superconductivity is found) are 1 and 20 $\mu\Omega$cm for Sr$_2$RuO$_4$ (Ref. \onlinecite{Mackenzie1998}) and CeCoIn$_5$,\cite{Petrovic2002,Paglione2007} respectively.

The discovery of high $T_c$ superconductivity in the AEFe$_2$As$_2$ refocused attention on Fe-bearing compounds with ThCr$_2$Si$_2$ structure, especially ones that manifest reduced moment magnetic ordering. The isostructural \Yott~(Ref. \onlinecite{Venturini1996,Avila2004}) was an obvious candidate for attention. The temperature dependent resistivity exhibits metallic behavior with the residual resistivity ratio of $\sim$30 along the $ab$-plane in single crystal samples grown out of Sn flux.\cite{Avila2004} Whereas it is lacking any transitions down to 2 K, it does exhibit a large Pauli paramagnetic susceptibility and the Sommerfeld coefficient of 3$\times$10$^{-3}$ emu/mol and $\gtrsim$ 60 mJ K$^{-2}$mol$^{-1}$, respectively.\cite{Avila2004} For comparison, the similar transition metal bearing compounds, YCo$_2$Ge$_2$ (Ref. \onlinecite{Kong2014}) and YNi$_2$Ge$_2$ (Ref. \onlinecite{Budko1999}) are typical metals. Interestingly, the lutetium variant of \Yott, \Luott, exhibits a transition at 10 K, which was attributed to antiferromagnetic transition due to Fermi surface nesting,\cite{Avila2004,Ferstl2006,Fujiwara2007} similarly with parent compounds of the Fe-based superconductors. Based on these observations Ran {\it et al.} performed extensive substitution studies on Lu$_{1-x}$Y$_x$Fe$_2$Ge$_2$ and found that Y-substitution for Lu suppresses the magnetic transition.\cite{Ran2011} The critical concentration, at which the antiferrormagnetic transition is suppressed to zero temperature, is $x\approx 0.20$.\cite{Ran2011} Although, in some cases, this is where a maximum $T_c$ is expected to be found,\cite{Varma2010} no superconductivity was observed in Lu$_{0.814}$Y$_{0.186}$Fe$_2$Ge$_2$ down to 2 K.\cite{Ran2011} In light of these studies, the recent preprint by Zou {\it et al.}\cite{Zou2013} presenting resistivity data on polycrystalline samples of \Yott~with a reported $T_c$ value of 1.8 K was of specific interest.

The Cambridge University group of Zou {\it et al.} reported the observation of abrupt changes in both electrical and magnetic properties at low temperatures in a polycrystalline sample of \Yott~fabricated via arc melting and a subsequent annealing at 800\celsius. The electrical resistivity drops to zero, and the sample  expels weak magnetic fields at low temperatures ($T<1.8$ K),\cite{Zou2013} which is consistent with the presence of a superconducting phase in this sample.\cite{Zou2013} {\it If} this superconductivity is found to be bulk and intrinsic to \Yott, then this stoichiometric compound could possibly be added into the class of Fe-based superconductors as a new member with its electron count being equivalent to the nodal superconductor KFe$_2$As$_2$,\cite{Singh2014} which is one of the rare examples of undoped superconductors among the Fe-based ones. However, there is, to date, no evidence for bulk superconductivity in \Yott.

Finding superconductivity in \Yott~is intriguing particularly because the compound has high density of states at the Fermi level as inferred from the large Sommerfeld coefficient\cite{Avila2004,Zou2013} and the electrical resistivity varies as $\rho(T)\propto T^{1.5}$, which is possibly due to proximity of this compound to a magnetic quantum critical point.\cite{Zou2013,Singh2014} Based on results of first-principle calculations, the superconductivity observed in this compound is compatible with either the sign-changing multi $s$-wave, so-called $s_\pm$-wave,\cite{Subedi2014} or spin fluctuation mediated spin triplet superconductivity,\cite{Singh2014} some of which might have the superconducting transition temperature depending sensitively on disorder. However, there has been no systematic work on the relation between the superconducting transition temperature and the impurity scattering in YFe$_2$Ge$_2$. Also, the maximum residual resistivity ratio in the single crystal samples remains about 30 (Refs. \onlinecite{Avila2004,Ran2011}) whereas the polycrystalline samples that manifest signatures of superconductivity in transport data have RRR values between 30 and 50.\cite{Zou2013} 

\begin{figure}
%\centering
\includegraphics[width=1\linewidth]{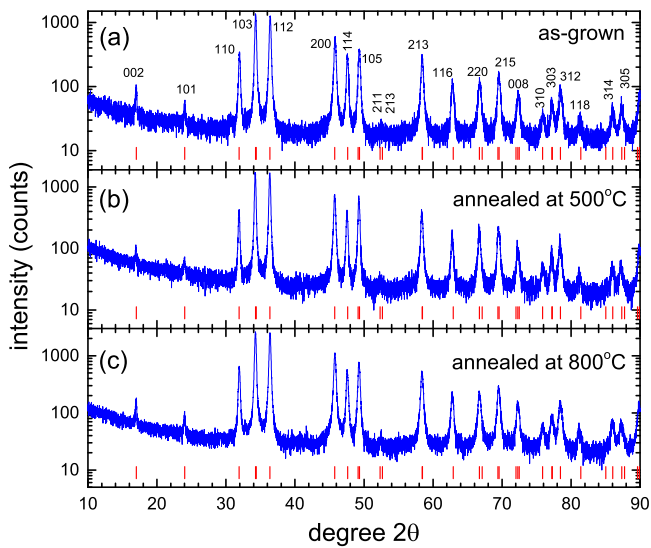}
\caption{Powder x-ray patterns of as-grown and annealed at 500\celsius~and 800\celsius~samples. Note: whereas the as-grown and 800\celsius-annealed samples do not have resistive signatures of superconductivity, the 500\celsius~annealed samples do have them (see text).}
\label{fig:xray}
\end{figure}

In this work we report on improving the residual resistivity ratio (RRR) of single crystalline \Yott~up to $\sim$60 by modifications in growth as well as by post growth annealing. We find that samples with RRR values greater than 34 manifest signatures of a superconducting transition in transport data, but improvements of RRR from 34 to 63 do not lead to any systematic changes in $T_c$. Specific heat was measured on samples which show relatively sharp, resistive superconducting transitions at temperature $\sim$1 K, but no anomaly associated with superconductivity was observed down to 0.4 K. The absence of features in specific heat suggests that the superconductivity observed in this compound is either of filamentary, strain stabilized superconductivity associated with small amounts of stressed YFe$_2$Ge$_2$, possibly at twin boundaries\cite{Khlyustikov1987,Logvenov2009} or dislocations or is a second crystallographic phase present at level below our detection capability.

%%%%%%%%%%%%%%%
%%%%%%%%%%%%%%%
%%%%%%%%%%%%%%%
\section{Experimental}

%\subsection{crystal growth}

Single crystals of \Yott~were grown out of Sn flux.\cite{Canfield1992,Avila2004,Ran2011} The elements were mixed together into a 2 ml Al$_2$O$_3$ crucible according to the ratio Y:Fe:Ge:Sn $\approx$ 1:2.4:2:95 with the mass of Sn being typically $\sim$5 g. The crucible with starting elements was sealed in a fused-silica ampule under a partial argon atmosphere. The ampoule was then placed in a box furnace. The elements were dissolved and mixed in molten Sn by holding the temperature at 1190\celsius~for 2 hours, then the crystals grew while  the melt cooled over at least four days to the decanting temperature which varied from 300\celsius~to 550\celsius, at which point the ampoule was quickly removed from the furnace, and the molten Sn flux was decanted using a centrifuge.\cite{Canfield1992} All samples were etched in concentrated HCl for about 30 min to remove residual Sn from the crystal surface before any measurements were done.

%\subsection{annealing}

For annealing, samples were placed into an alumina combustion boat. The boat was then placed into a fused-silica tube that was continuously pumped on by a turbo-molecular pump maintaining pressures lower than 10$^{-6}$ torr throughout the annealing time. The tube containing the alumina boat and samples was heated up to a desired temperature between 400\celsius~and 800\celsius~for a desired period of time, typically a week.

%\subsection{powder x-ray and Laue}

Powder x-ray diffraction measurements were performed at room temperature using a Rigaku Miniflex diffractometer with Cu $K\alpha$ radiation. Diffraction patterns were taken on ground single crystals of as-grown samples decanted at 550\celsius~and samples annealed at temperatures 500\celsius~and 800\celsius. All lines can be indexed to the reported \Yott~structure, and no extra lines appear as a result of annealing as shown in Fig. \ref{fig:xray}.

\begin{figure}
%\centering
\includegraphics[width=0.7\linewidth]{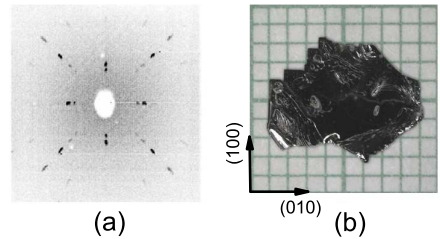}
\caption{(a) x-ray Laue backscattering pattern for YFe$_2$Ge$_2$, showing a four-fold rotation symmetry of the [001] direction. Horizontal axis is a [100] direction. (b) A sample that was used to collect the Laue-back-reflection pattern.}
\label{fig:laue}
\end{figure}

The x-ray Laue-back-reflection pattern was taken with a MWL-110 camera manufactured by Multiwire Laboratories. Figure \ref{fig:laue} presents an x-ray Laue backscattering pattern and a single crystal of YFe$_2$Ge$_2$ used to collect this pattern respectively. Figure \ref{fig:laue}(a) shows a four-fold rotation symmetry and four mirror planes, with 45\degree~angles between each other, of the $c$-axis. The [100] directions are along the naturally formed edges of the single crystal shown in Fig. \ref{fig:laue}(b).

%\subsection{electrical transport}

For in-plane electrical transport measurements, rectangular samples were cut with a wire-saw out of plate-like single crystals. Once cut, the samples were cleaned using acetone and ethanol. If needed, the sample-surfaces were polished by sanding on a silicon carbide paper with 4000 grit. Electrical contacts to the samples were made using Epotek H20e silver epoxy and platinum wires in a standard four-probe geometry. The epoxy was cured for approximately 30 min at 120\celsius. The typical contact resistance is about 0.5 - 1.0 $\Omega$ at each contact. An {\it Oxford Instrument} dilution refrigerator, a {\it CRYO Industry of America, Inc.} $^3$He cryostat and a {\it Quantum Design Physical Property Measurement System} (PPMS) were employed to measure in-plane resistivity over the temperature range between 50 mK and 305 K. For the PPMS, its internal resistance bridge was used with the AC transport option. For the other cryostats, a {\it Lake Shore} AC resistance bridge model 370 (LS370) was used. The small mismatch between data taken by the PPMS internal bridge and LS370 was corrected by vertically shifting the LS370-data to match with the PPMS-data.
 
 %\subsection{heat capacity}
 
Specific heat measurements were performed in a {\it Quantum Design PPMS} with a $^3$He cooling option allowing measurements down to 0.4 K, using the relaxation calorimetry technique. Before measuring specific of a sample, background data (addenda) were taken with a small amount of {\it Apiezon N} grease which was subsequently used to mount the sample.

%%%%%%%%%%%%%%%
%%%%%%%%%%%%%%%
%%%%%%%%%%%%%%%
\section{Results and discussion}

%%%%%%
%%%%%%
\subsection{RRR vs. growth details}

% Figure RRR vs T
\begin{figure}
\includegraphics[width=0.8\linewidth]{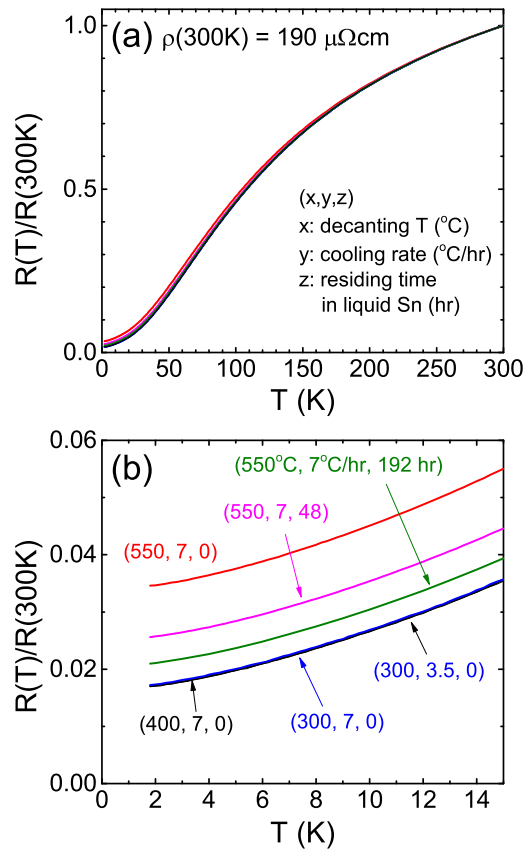}%
\caption{RRR vs. growth details. (a) Normalized resistance $R(T)/R$(300K) vs. $T$ of samples grown by various conditions, showing an almost identical temperature variation over wide temperature ranges. The growth conditions are represented by $(x,y,z)$ where $x$, $y$, and $z$ are the decant temperature, cooling rate, and the time for which crystals resided in molten Sn at the desired decanting temperature until the actual decanting. (b) Enlarged, low temperature data from curves shown in the upper panel.}
\label{fig:rrr}
\end{figure} 

In order to maximize the residual resistivity ratio, RRR, several parameters were tuned in the growth profile. The tuning parameters include the decanting temperature, cooling rate, and additional residing time in molten Sn after a desired decanting temperature is reached (annealing in solution). Figure \ref{fig:rrr} shows normalized resistances $R(T)/R$(300K) of the six representative samples grown with various conditions. The resistance measurements were made on unpolished samples. Panel (a) shows the data on a full temperature range. All samples behave roughly the same in this scale. However, small deviations start developing around 150 K upon cooling. The effects of the different growth conditions are more obvious at low temperatures as displayed in Fig. \ref{fig:rrr}(b). The sample, cooled at a rate of 7\celsius/hr and immediately decanted at 550\celsius, shows the smallest $R$(300K)/$R$(1.8K)$\equiv$RRR $\approx$ 29. RRR can be enhanced by keeping single crystals in molten Sn before decanting. RRR values of 39 and 48 can be achieved with wait-time of 48 and 192 hours, respectively, before decanting. Alternatively, even higher RRR can be achieved by decanting at lower temperatures. Cooling samples in molten Sn further down to 300\celsius~or 400\celsius~at the same rate increases RRR to $\sim$60, which is almost twice as large as the previously reported values in the single crystals.\cite{Avila2004,Ran2011,Zou2013} On the other hand, slower cooling with the rate of 3.5\celsius/hr to 300\celsius~does not further improve the RRR.

% Figure RRR vs growth
\begin{figure}
\includegraphics[width=0.7\linewidth]{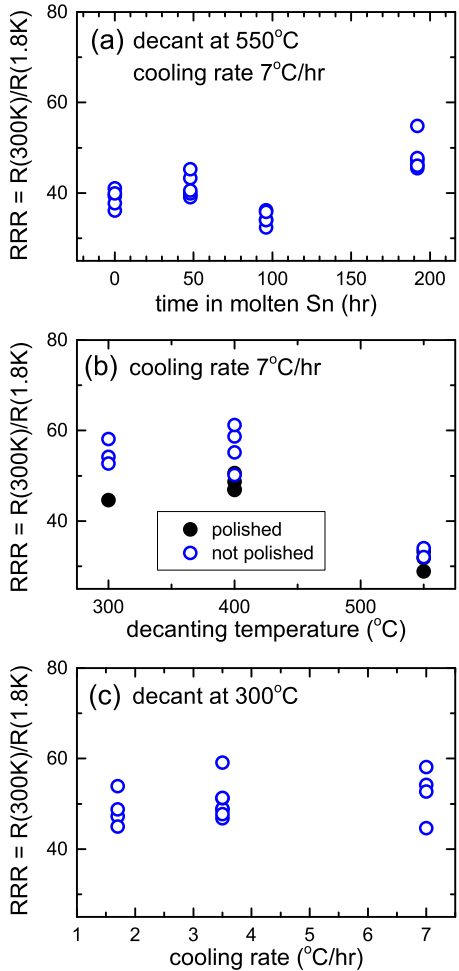}%
\caption{RRR vs. growth details. (a) Residing time in molten Sn after cooling to 550\celsius~with a fixed cooling rate of 7\celsius/hr. (b) Decanting temperature with a fixed cooling rate of 7\celsius/hr without any additional time in molten Sn. Open and solid symbols correspond to unpolished and polished samples, respectively. (c) Cooling rate to the immediate-decanting $T=300$\celsius.}
\label{fig:growth}
\end{figure} 

For more statistics, 4 to 7 samples were measured from each batch, and the results are summarized in Fig. \ref{fig:growth}. Fig. \ref{fig:growth}(a) shows RRR of the samples which were kept in molten Sn for a desired time at 550\celsius~before decanting. Other than variation of the residing time in molten Sn, the growth profile is identical to that used in Refs. \onlinecite{Avila2004,Ran2011}. The average RRR values are $39$, $41$, $35$, and $48$ for 0, 48, 92 and 192 hours, respectively. Whereas an 192 hr dwelling time in molten Sn slightly improves the RRR value, there is, at best, a weak dependence of RRR on this parameter.

Next, we changed the decanting temperature while other parameters were fixed. Figure \ref{fig:growth}(b) shows RRR as a function of decanting temperature. Open and solid symbols represent RRR in unpolished and polished samples, respectively. The average RRR values are $52$, $53$, and $32$ for decanting temperatures of 300\celsius, 400\celsius~and 550\celsius, respectively. This result clearly shows that decanting at 300\celsius~and 400\celsius~is better than decanting at 550\celsius. It is noteworthy that RRR in polished samples shows somewhat smaller values in all batches tested here.

Lastly, we changed the cooling rate with a fixed decanting temperature of 300\celsius. Figure \ref{fig:growth}(c) shows RRR versus various cooling rates of samples from three different batches decanted immediately at 300\celsius. The average RRR values are $52$, $51$, and $48$ for the batches with cooling rates of 7\celsius/hr, 3.5\celsius/hr and 1.7\celsius/hr, respectively. There is practically no effect of slowing the cooling rate on RRR within error bars over the RRR-ranges tested. We would like to note, though, that the thickness of plate-like single crystals increases up to 0.4 mm with the slowing of the cooling rate, which may allow direct transport measurements with current along the $c$-direction.

From these experiments, we may conclude that decanting at temperatures between 300\celsius~and 400\celsius~with any cooling rate between 7\celsius/hr and 1.7\celsius/hr gives the largest RRR values.

%%%%%%%%%%%%%%%
%%%%%%%%%%%%%%%
\subsection{Sample polishing effect on RRR}\label{sec:polishing}

% Figure RRR vs growth details
\begin{figure}
\includegraphics[width=0.9\linewidth]{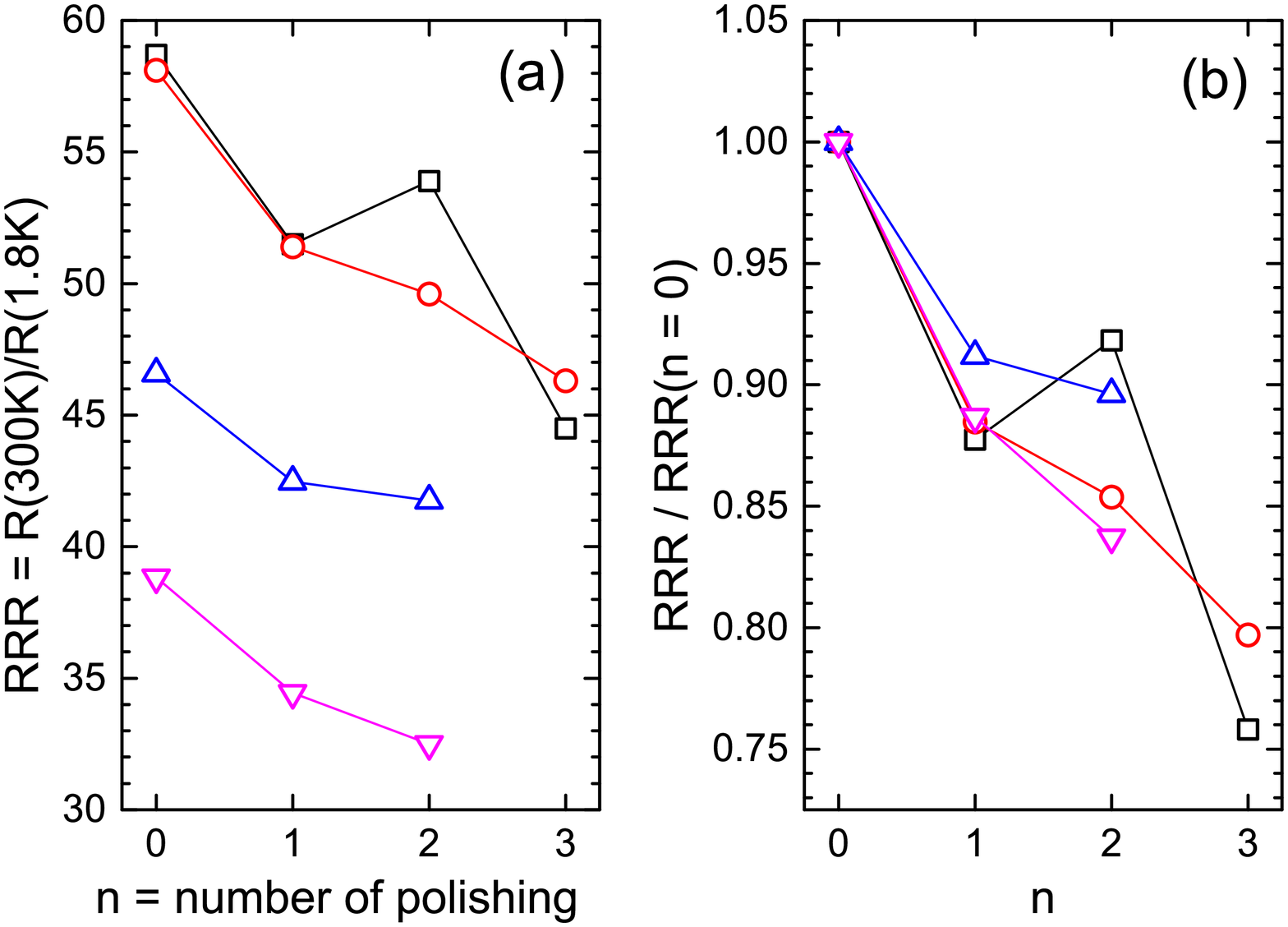}%
\caption{Effect of polishing. (a) Change in RRR of as-grown samples after polishing. (b) The normalized values by the initial RRR values of the pristine samples.}
\label{fig:polishing}
\end{figure} 

Noting that we found smaller RRR in polished samples (Fig. \ref{fig:growth}b), we examined the effect of polishing on the samples by measuring RRR before and after polishing. Four samples with different initial RRR varying between 38 and 60 were selected. After the first resistivity measurements on as-grown samples, the samples were polished into thin, long bars. The following experiments were done with the new contacts on freshly polished surfaces. In each cycle, the samples underwent a brief heating at 120\celsius~for typically 30 minutes in addition to polishing to attach contacts (see Experimental methods). Effects of this 120\celsius~curing were checked separately by re-measuring a resistivity-bar with contacts after annealing at the same temperature for 30 minutes, and there were no detectable effects on the resistivity measurements.

The results of sequential polishing on six samples are summarized in Fig. \ref{fig:polishing}. Regardless of initial RRR, the first polishing reduces RRR. The second polishing also reduces RRR except for one sample that is represented by square symbols. A third polishing for two samples reduces RRR to about 75\% of the initial RRR values. The normalized scale shown in Fig. \ref{fig:polishing}(b) reveals that RRR is reduced by $\sim$10\% by initial polishing and further (up to 25\%)  by subsequent polishing. This result clearly demonstrates that these samples are relatively easily degraded by light mechanical work, likely forming dislocations or similar defects.

%The two samples with high RRR were annealed, and RRR after annealing is xxx.

%%%%%%%%%%%%%%%
%%%%%%%%%%%%%%%
\subsection{Annealing effect}

% Figure RRR vs Annealing
\begin{figure}
\includegraphics[width=0.9\linewidth]{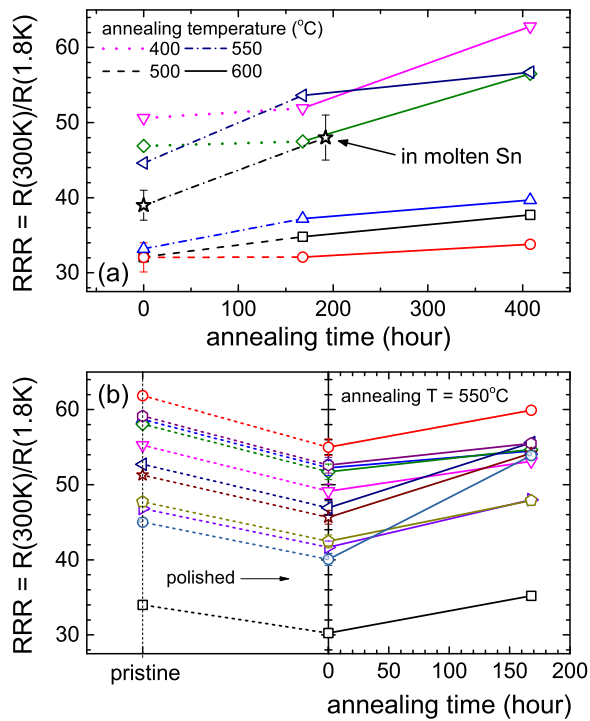}%
\caption{Evolution of RRR upon annealing. (a) Change in RRR by annealing with contacts for 168 hours at $T=$ 400\celsius, 500\celsius, and 550\celsius, represented by dotted, dashed, dot-dashed lines, followed by a subsequent annealing at 600\celsius~for additional 240 hours, indicated as solid lines. For comparison the average RRR of the samples annealed in molten Sn of temperature 550\celsius~is added to the plot (starred symbol). (b) Another set of samples were measured as as-grown first, followed by polishing and annealing without electrical contacts at 550\celsius~for 168 hours. The intermediate RRR values are estimated by applying 11\% reduction on initial RRR values in unpolished, pristine samples.}
\label{fig:annealing}
\end{figure} 

To investigate the effects of post growth annealing, two sets of samples were prepared with initial RRR values varying between 30 and 60. The results of the resistivity measurements done on the first set of samples are displayed in Fig. \ref{fig:annealing}(a) (annealing time = 0). These samples were cut and polished before these initial resistivity measurements. They were subsequently annealed together with the contacts which include platinum wire and a small amount of silver epoxy. First, three pairs of samples were annealed at three different temperatures of 400\celsius, 500\celsius, and 550\celsius~for 168 hours, and the change in RRR is indicated in dotted, dashed, and dashed-dot lines, respectively. Whereas annealing at 400\celsius~(dotted lines) is not so effective, annealing at higher temperatures shows clear enhancement in RRR. For comparison the average RRR of the samples annealed in molten Sn of temperature 550\celsius~for 192 hours is also shown on the plot (starred symbol), which shows almost the same effect as 550\celsius-annealing of the samples with the contacts. Subsequently, all samples were annealed at 600\celsius~for additional 240 hours, and the result is shown in solid lines in Fig. \ref{fig:annealing}(a). All samples show enhanced RRR, although the effect is rather small for the samples with initial RRR less than 35.

Although presence of silver and platinum for contacts does not seem to affect RRR, judging from the similar results with annealing in molten Sn, we investigated annealing effect for another set of samples which were annealed without contacts. Eleven samples were selected from various batches, and resistivity measurements were made on as-grown, unpolished samples. After the first resistivity measurements, each sample was polished into a bar shape. Subsequently, they were annealed at 550\celsius~for 168 hours. RRR of each sample after polishing was estimated by applying the approximate reduction rate of $11\pm2$ \% determined in Section \ref{sec:polishing}. The evolution of RRR during this experiment is presented in Fig. \ref{fig:annealing}(b). There is some sample dependence, but overall the results are similar to the ones shown in Fig. \ref{fig:annealing}(a). More importantly, though, when we compare the RRR vales of the pristine samples to those of the polished and annealed ones we find very little overall change. It becomes clear that the annealing is most likely repairing the damage introduced during polishing, provided the estimated reduction in RRR is applicable for these samples. 

In addition, several samples were annealed at 700\celsius~and 800\celsius~for two days, and we found that RRR of these samples decreased to $\sim$20.

%%%%%%%%%%%%%%%
%%%%%%%%%%%%%%%
\subsection{Superconductivity}
\subsubsection{Resistivity measurements}
%\subsubsection{appearance of superconductivity by annealing}

% Figure TcvsRRR SAxxx samples
\begin{figure}
\includegraphics[width=0.9\linewidth]{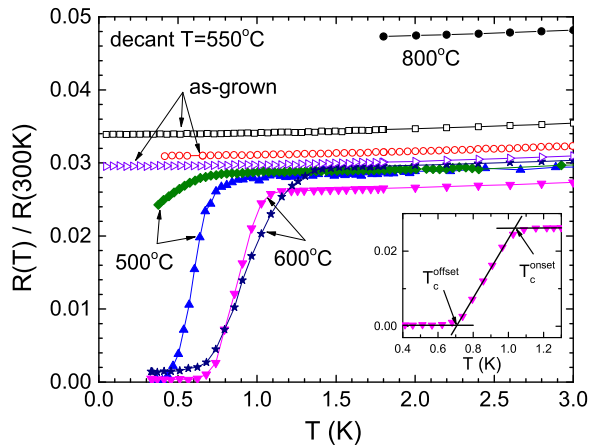}%
\caption{$R(T)/R$(300K) for representative as-grown (decanted at 550\celsius) and annealed samples shown in open and solid symbols, respectively.}
\label{fig:resisSA}
\end{figure} 

A representative set of $R(T)/R$(300K) data ($T<2.5$ K) of as-grown and annealed samples data are presented in Fig. \ref{fig:resisSA}.  All samples displayed in the figure are obtained by immediate decanting at $T=550$\celsius. A typical RRR of as-grown samples is about 30 with a small variation. As-grown samples do not exhibit any signature of superconductivity down to 50 mK. Annealing at temperatures of 500\celsius~and 600\celsius~for 7 and 10 days, respectively, induces a superconducting transition, and annealing at the higher temperatures is more effective. Representative superconducting transition temperatures, $T_c$, induced by annealing at 500\celsius~and 600\celsius~are 0.6 K and 0.9 K shown as upward-and downward-triangles, respectively, shown in in Fig. \ref{fig:resisSA}. Here $T_c$ is defined at the midpoint of $T_c^\textmd{\tiny onset}$ and $T_c^\textmd{\tiny offset}$, {\it i.e.}, $T_c=(T_c^\textmd{\tiny onset} + T_c^\textmd{\tiny offset})/2$ (see the inset of Fig. \ref{fig:resisSA}). However, as discussed above, annealing at 800\celsius~for a week is apparently disadvantageous because RRR decreased to $\sim$20.

%\subsubsection{appearance of superconductivity in as-grown samples}

% Figure TcvsRRR XHxxx samples
\begin{figure}
\includegraphics[width=0.9\linewidth]{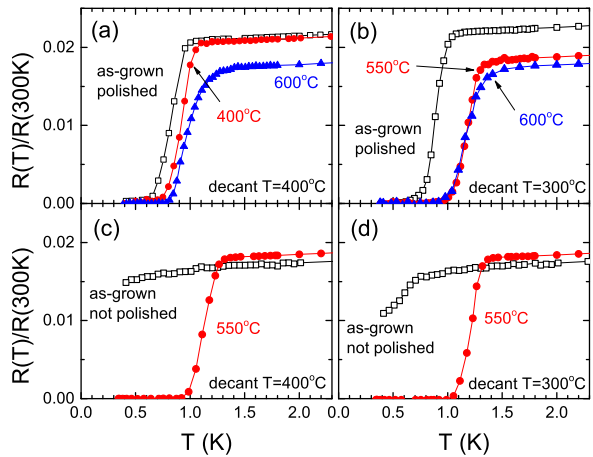}%
\caption{$R(T)/R$(300K) in as-grown and annealed samples represented by open and solid symbols, respectively. Samples shown in (a),(c) and (b),(d) are from two batches which were decanted at 400\celsius~and 300\celsius, respectively.}
\label{fig:resisXH}
\end{figure} 

Figure \ref{fig:resisXH} shows $R(T)/R$(300K) of the samples which are decanted at 400\celsius~and 300\celsius~in (a), (c) and (b), (d), respectively. Unlike the batch decanted at 550\celsius, the samples in these batches all show a signature of superconducting transition with greatly enhanced RRR. Upper panels (a) and (b) show the normalized resistance in as-grown, polished samples in open symbols, which show a complete transition with $T_c\approx 0.75$ K for both samples. $T_c$ can be increased by annealing at temperatures between 400\celsius~and 600\celsius~accompanied by increase of RRR values as shown in solid symbols.

On the other hand, some as-grown, unpolished samples do not show a full transition as shown in Figs. \ref{fig:resisXH}(c) and \ref{fig:resisXH}(d) even though they exhibited much greater RRR. The relation between polishing and full-transition is not clear at the moment although this may indicate that superconductivity is an artifact associated with damaged or strained surfaces, in some manner similar to the case of spurious superconductivity in ZrZn$_2$.\cite{Yelland2005} After the first resistivity measurements on unpolished surface, these two samples were polished into a bar-shape and subsequently annealed at 550\celsius~for a week. After annealing, the sharp superconducting transition is evident at $T_c=1.1$ K, but with slightly decreased RRR values for both samples.

%\subsubsection{$T_c$ vs RRR}

% Figure TcvsRRR
\begin{figure}
\includegraphics[width=0.9\linewidth]{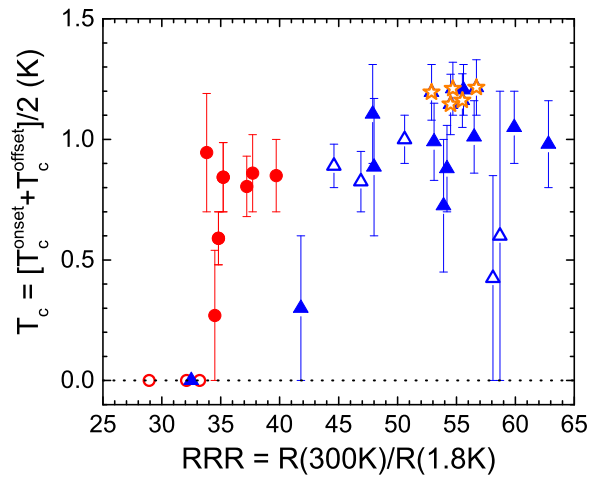}%
\caption{$T_c$ vs. $R$(300K)/$R$(1.8K). Open and solid symbols represent as-grown and annealed samples, respectively. The circles represent samples which are not superconducting in as-grown form and the triangles represent samples which show signature of superconducting transition in as-grown form. The five samples with RRR$\approx$55 marked by open stars were used for heat capacity measurements described in Section (\ref{sec:Cp}).}
\label{fig:TcvsRRR}
\end{figure}

Figure \ref{fig:TcvsRRR} summaries the relation of $T_c$ and RRR for all samples studied. Open and solid symbols represent as-grown and annealed samples, respectively, and the length of the vertical error bars is the width of superconducting transition, {\it i.e.}, temperature difference between $T_c^\textmd{\tiny onset}$ and $T_c^\textmd{\tiny offset}$. The red, circular symbols represent the batches decanted at 550\celsius, and the as-grown samples in the batch do not show any signature of superconductivity. The superconductivity observed in the samples from this batch was achieved by annealing at temperatures of 500\celsius~and 600\celsius. The blue triangles represent the batches in which as-grown samples show superconductivity as indicated in open triangles. Overall, superconductivity appears in samples with RRR $>$ 34, and annealing at temperature $\sim$550\celsius~gives $T_c$ values up to 1.2 K.

It should be noted, though, that beyond and apparent minimum RRR value, there is essentially no effect of RRR on $T_c$, {\it i.e.}, $T_c$ is independent of RRR for a range between 34 and 63.

%\subsubsection{fragile superconductivity}

% Figure resistivity comparison
\begin{figure}
\includegraphics[width=0.9\linewidth]{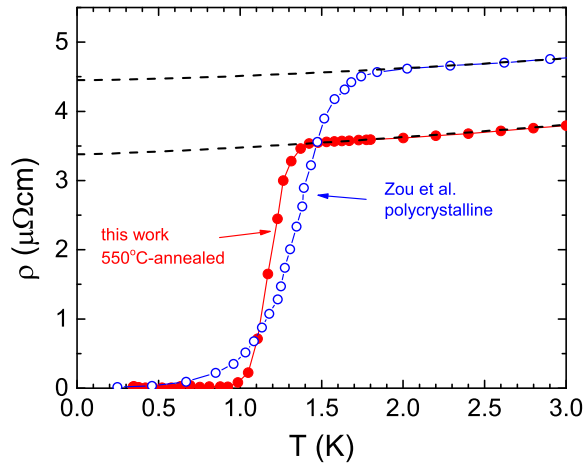}%
\caption{Temperature dependence of resistivity, $\rho(T)$, at low temperatures ($T<3$ K). The data shown by solid circles are from one of samples annealed at 550\celsius. The data shown by open circles are taken from Ref. \onlinecite{Zou2013} for comparison. The dashed lines are determined by fitting the data to a power-law function, $\rho(T)=\rho_0+AT^n$ with $\rho_0$, $A$, and $n$ being fitting parameters.}
\label{fig:rhocomp}
\end{figure} 

In Figure \ref{fig:rhocomp}, we compare $\rho(T)$ data for one of our crystals to a polycrystalline sample from Ref. \onlinecite{Zou2013}. The polycrystalline sample\cite{Zou2013} shows greater $\rho_0$ than the single crystal sample annealed at 550\celsius~and a higher $T_c^\textmd{\tiny onset}$ with a significantly broader width of transition leading to a lower $T_c^\textmd{\tiny offset}$. $\rho_0$ was determined by extrapolating normal state $\rho(T)$ data below 10 K using a power-law function, $\rho(T)=\rho_0+AT^n$, and the determined $\rho_0$ is 3.38 and 4.45 $\mu\Omega$cm for the single crystal sample and the polycrystalline sample, respectively. The values of ($n$, $A$) obtained from the fitting procedure are (1.4, 0.10 $\mu\Omega$cm/K$^{1.4}$) and (1.5, 0.06 $\mu\Omega$cm/K$^{1.5}$) for the single crystal and polycrystalline samples, respectively.

It is noteworthy that the temperature variation of in-plane resistivity of single crystalline samples in a temperature range below 10 K is $\Delta\rho(T)\propto T^n$ with $n$ varying between 1.4 and 1.6, similarly with the observed value in the polycrystalline sample.\cite{Zou2013}

%%%%%%%%%%%%%%%
%%%%%%%%%%%%%%%
\subsubsection{Specific heat measurements}\label{sec:Cp}

\begin{figure}
\includegraphics[width=0.9\linewidth]{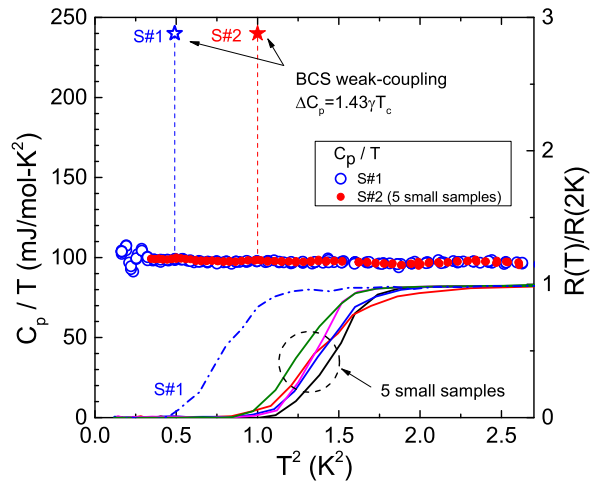}%
\caption{Specific heat and resistivity in YFe$_2$Ge$_2$. S\#1 represents measurements on a sample with mass 6 mg shown in open circles and dashed-dot lines for $C_p/T$ and $R(T)/R$(2K), respectively. S\#2 represents measurements done on a collection of 5 small samples with total mass of 2.5 mg shown in solid circle and lines, for $C_p/T$ and $R(T)/R$(2K), respectively. The open and solid starred symbols represent BCS weak-coupling predictions of a specific heat jump for S\#1 and S\#2, respectively.}
\label{fig:Cp}
\end{figure}

In order to test for bulk rather than filamentary or second phase superconductivity, temperature dependent specific heat $C_p$ was measured first on a sample (S\#1) with mass of 6 mg annealed at 550\celsius~for 7 days. Low temperature $C_p/T$ of this sample is shown in blue, open circles in Fig. \ref{fig:Cp}. It has weak temperature dependence which is slowly increasing upon cooling, and is consistent with the data taken on a polycrystalline sample.\cite{Zou2013} The data of the $C_p(T)$ at higher temperatures show excellent agreement with the results presented in Ref. \onlinecite{Avila2004} (data not shown). This implies that the annealing does not affect normal state properties. Sommerfeld coefficient was determined to be 100 mJ/mol$\cdot$K$^2$, which is consistent with that reported in Refs. \onlinecite{Avila2004,Zou2013}.  The resistive superconducting transition in this sample was checked after $C_p$ measurement was done, and the normalized resistance data, $R(T)/R$(2K), at low temperatures with $T^\textmd{\tiny offset}_c\approx 0.7$ K ($0.49$ K$^2$ in Fig. \ref{fig:Cp}) is shown in the right vertical axis in Fig. \ref{fig:Cp}. An expected specific heat jump, $\Delta C_p$, estimated in the BCS weak-coupling limit, $\Delta C_p=1.43\gamma T_c$,\cite{Tinkham1996} is marked in a blue, open star.

Another specific heat measurement (S\#2) was done on five selected samples with a relatively sharp transition in their resistivity measurements, marked by open stars in Fig. \ref{fig:TcvsRRR}. Electrical contacts were carefully removed before the specific heat measurement. Total mass of the samples was 2.5 mg, including a small amount of cured silver epoxy from the contacts. $C_p/T$ data of the samples at low temperatures are shown in Fig. \ref{fig:Cp} along with the normalized resistivity curves (solid lines) for each sample on the right vertical axis. Although these samples have sharper transitions than that shown by Zou {\it et al.},\cite{Zou2013} the $C_p(T)$ data display no apparent anomaly across the temperature region where the resistivity data show clear superconducting transitions. A BCS prediction is also marked in a red, solid star by using the same BCS relation.

Both measurements suggest that the superconducting transition observed in resistivity is not bulk but is either associated with a  filamentary second phase or some sort of surface strain stabilized superconductivity. Given the fact that our x-ray results on superconducting and non-superconducting samples are practically identical within detection limit, formation of a secondary phase in the twin boundary which is typically sub-micron thick,\cite{Soni1995} could be responsible for the resistive superconducting transition.

% Figure Cp

%%%%%%%%%%%%%%%
%%%%%%%%%%%%%%%
%%%%%%%%%%%%%%%
\section{Summary and conclusions}

We report on improving the residual resistivity ratio (RRR) of \Yott~grown out of Sn up to 60 by tuning growth profile as well as by post growth annealing. We found decanting molten Sn at temperatures between 300\celsius~and 400\celsius~is very effective in enhancing RRR up to $\sim$60. In addition, post growth annealing at temperature $\sim$550\celsius~enhances RRR up to 60, as well. However, as-grown samples show a strong dependence on polishing, and reduction of RRR upon polishing was determined to be approximately 10\% in as-grown samples. We observed a resistive superconducting transition in samples with RRR greater than 34 and residual resistivity less than 5 $\mu\Omega$cm. The $T_c$ was found to be as high as 1.2 K with a relatively sharp transition compared to the transition in a polycrystalline sample, but the onset of the transition is somewhat lower in the single crystal sample. However, improvements of RRR from 34 to 60 do not lead to any systematic changes in $T_c$. Specific heat was measured in samples which show relatively sharp, resistive superconducting transitions at temperature $\sim$1 K, but no anomaly associated with superconductivity was observed down to 0.4 K. The absence of features in specific heat suggests that the superconductivity observed in this compound is either of filamentary, strain stabilized superconductivity associated with small amounts of stressed YFe$_2$Ge$_2$ possibly at twin boundaries or dislocations or is a second crystallographic phase present at level below our detection capability.

%%%%%%%%%%%%%%%
%%%%%%%%%%%%%%%
%%%%%%%%%%%%%%%
\section{Acknowledgements}
H.K. would like to thank V. Taufour for useful discussions. This work was carried out at the Iowa State University and supported by the AFOSR-MURI grant No. FA9550-09-1-0603 (H.K., E.D.M. and P.C.C.). Part of this work was performed at Ames Laboratory, US DOE, under contract No. DE-AC02-07CH 11358 (S.R., H.H., M.A.T., R.P. and S.L.B.).

%%%%%%%%%%%%%%%
%%%%%%%%%%%%%%%
%%%%%%%%%%%%%%%
%

\end{document}